\documentclass[preprint2]{aastex}

\usepackage{lscape}
\usepackage{apjfonts}

\newcommand{\footstar}[1]{$^*$ \footnotetext{$^*$#1}}

\shorttitle{25 Sub-Arcsecond Binaries}

\shortauthors{Richichi et al.}

\begin{document}


\title{Twenty-Five Sub-Arcsecond Binaries Discovered By Lunar Occultations
\footstar{Based on observations made with ESO telescopes at Paranal Observatory}
}


\author{A. Richichi\altaffilmark{1}, O. Fors\altaffilmark{2,3}, 
F. Cusano\altaffilmark{4}, and M. Moerchen\altaffilmark{5,6}}

\altaffiltext{1}{National Astronomical Research Institute of Thailand, 191 Siriphanich Bldg., 
Huay Kaew Rd., Suthep, Muang, Chiang Mai  50200, Thailand}
\email{andrea4work@gmail.com}
\altaffiltext{2}{Departament Astronomia i Meteorologia and Institut de Ci\`encies del Cosmos (ICC), 
Universitat de Barcelona (UB/IEEC), Mart\'{\i} i Franqu\'es 1, 08028 Barcelona, Spain}
\altaffiltext{3}{Observatori Fabra, Cam\'{\i} de l'Observatori s/n, 08035, Barcelona, Spain}
\altaffiltext{4}{INAF-Osservatorio Astronomico di Bologna, Via Ranzani 1, 40127 Bologna, Italy}
\altaffiltext{5}{European Southern Observatory, Casilla 19001, Santiago 19, Chile}
\altaffiltext{6}{Space Telescope Science Institute, 3700 San Martin Drive, Baltimore, MD 21218, USA}



\begin{abstract}
We report on 25 sub-arcsecond binaries, detected for the first time 
by means of lunar occultations
in the near-infrared as part of a long-term program
using the ISAAC instrument at the ESO Very Large Telescope.
The primaries have magnitudes in the range
K=3.8 to 10.4, and the companions in the range
K=6.4 to 12.1. The magnitude differences have a median
value of 2.8, with the largest being 5.4.
The projected separations are in the range 6 to 748 milliarcseconds and
with a median of 18 milliarcseconds, or about 3 times less than
the diffraction limit of the telescope.
Among our binary detections are a pre-main sequence star
and an enigmatic Mira-like variable previously suspected to have
a companion. Additionally, we quote an accurate first-time near-IR
detection of a previously known wider binary.

We discuss our findings on an individual basis as far as made
possible by the available literature, and we
examine them from a statistical point of view. We derive
a typical frequency of binarity among field stars of $\approx 10$\%, in the
resolution and sensitivity range afforded by the technique 
($\approx 0\farcs003$ to  $\approx 0\farcs5$, and K$\approx 12$\,mag, respectively). 
This is in line
with previous results by the same technique but we point out interesting differences
that we can trace up to sensitivity, time sampling, and average
distance of the targets. Finally,
we discuss the prospects for further follow-up studies.
\end{abstract}





\section{Introduction}
\label{section:introduction}

Lunar occultations (LOs) can efficiently yield high angular
resolution observations from the analysis of the diffraction
light curves generated when background sources are covered
by the lunar limb. The technique has been employed
to measure  hundreds of stellar
angular diameters, binary stars, and sources with extended
circumstellar emission \citep[see CHARM2 catalog,][]{CHARM2}.
In the past few years, a program to observe LOs in the
near-infrared at the ESO Very Large Telescope (VLT) has
been very successful both in quantity, with over a thousand
events recorded, and in quality with a combination of
 angular resolution far exceeding the diffraction limit of a
single telescope ($\approx 0\farcs001$)
and a sensitivity significantly better than that
currently achieved by long-baseline interferometry
(K $\approx 12$\,mag). The drawbacks are that LOs are
fixed-time events, yielding mainly a one-dimensional scan
of the source, and that the source cannot be chosen at will.
Details on the LOs program at the VLT can be found
in \citet{apjssur}, and references therein.

Here, we report on 25 sources discovered to be binary
with projected separations below one arcsecond, and
in fact mostly below the 57 milliarcseconds (mas)
diffraction limit of the telescope at the given wavelength.
We also report on one previously known system.
In Sect.~\ref{section:data} we describe the
observational procedure, the sample composition, and the data analysis.
In Sect.~\ref{section:results} we report the
individual results, and provide some context
from previous bibliography when available.
Some considerations on the statistics of
binary detections from our VLT LO program
and on the prospects of follow-up of selected
systems are given in Sect.~\ref{sec:conclusions}.

\section{Observations and data analysis}
\label{section:data}

The observations were carried out between
April 2010 and October 2011, using
the 8.2-m UT3 Melipal telescope of
the VLT and the ISAAC instrument operated in burst mode.
Most of the observations were carried out in service mode,
based on a strategy of profiting from
short slots that might become available depending
on the atmospheric conditions and execution status
of other service programs. Consequently, the targets were
inherently random. A few observations were
parts of isolated nights dedicated to LO observations
in visitor mode, and in this case the sources were
selected on the basis of their colors and brightness
in very extincted regions of the Galactic Bulge.
The sources observed in service mode have the field
"Our ID" in Table~\ref{tab:list} beginning with 
P85 to P88, which are the ESO Periods under
consideration. The sources without this prefix
were observed in visitor mode.

\begin{figure}
\plotone{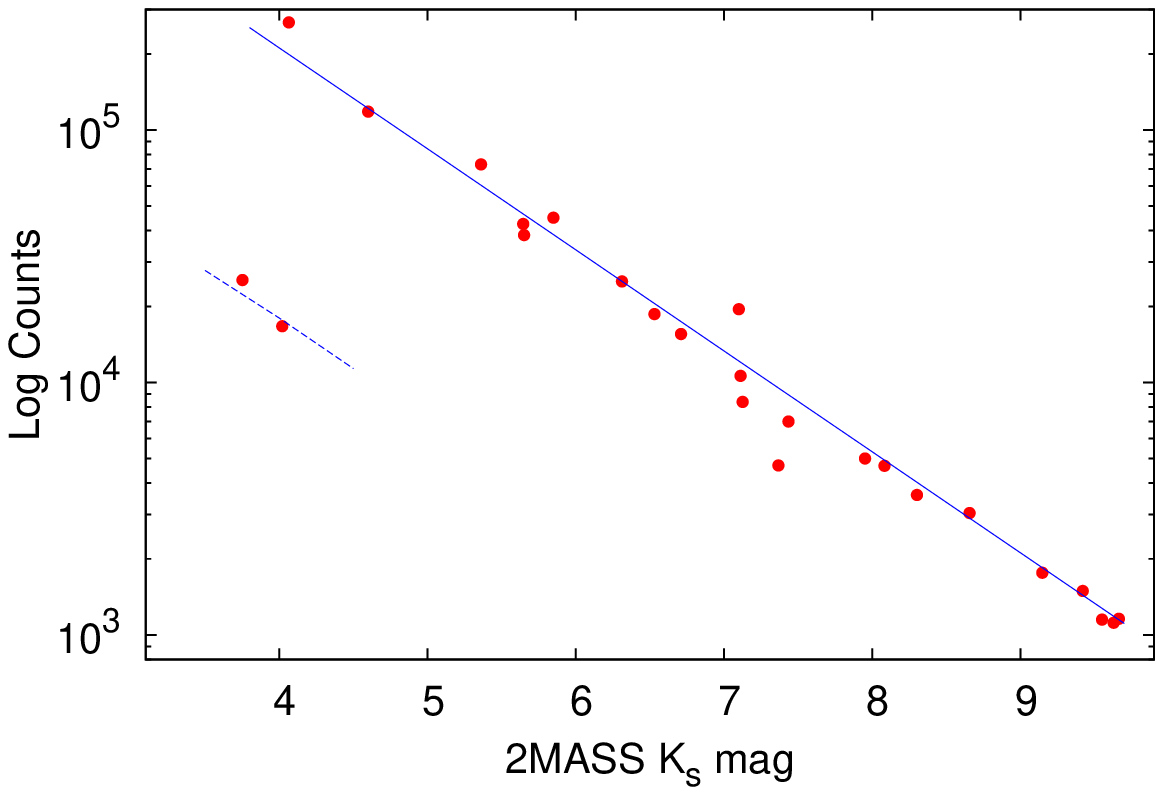}
\caption{Recorded fluxes (sum of the two components) as a function of
the 2MASS K magnitudes. The lines are the
expected fluxes based on the ISAAC Exposure Time Calculator,
for broad K$_{\rm s}$ and narrow filters (solid and dashed,
respectively).
}
\label{fig:counts}
\end{figure}

Table~\ref{tab:list} provides a 
list of the observations and of the
characteristics of the sources, ordered by time.
A sequential number is included, for ease
of cross-reference.
Our predictions were generated from
the 2MASS Catalogue, and so is the listed near-infrared
photometry. We did not attempt to derive 
proper K-band photometry from our light curves, due to
the lack of calibration sources.
However we notice that in some cases differences
between our counts and the 2MASS magnitudes
of up to a factor 2 are present, pointing to
possible variability (Fig.~\ref{fig:counts}).
Further
identifications in Table~\ref{tab:list}, 
as well as visual photometry and
spectral types, are
extracted from the {\it Simbad} database.

Each observation consisted of 7500 or 5000 frames
(in service or visitor runs, respectively)
in a 32x32-pixel ($4\farcs7 \times 4\farcs7$) sub-window,
with a time sampling of 3.2\,ms. This was also the effective
integration time. A broad-band K$_{\rm s}$ filter was
employed for all events, except in the case of
the brighter sources Stars 4 and 22,
for which
a narrow-band filter centered at 2.07\,$\mu$m was
employed to avoid possible non-linearities. 
The events were disappearances, with lunar
phases ranging from 38\% to 96\% (median 66\%). 
Airmass ranged  
from 1.1 to 2.0, while seeing ranged
from  $0\farcs6$ to $2\farcs9$ (median $1\farcs1$). 
The LO light curves are generated in vacuum at the
lunar limb and the diffraction fringes span a
range of few 0.1\,s, so that the technique
is in any case largely insensitive to atmospheric
perturbations.

The data cubes were converted to light curves by
extracting the signal at each frame within a
mask tailored to the seeing and image quality.
The light curves were then
analyzed using both a
model-dependent \citep{richichi96}
and a model-independent method (CAL, \citealt{CAL}).
This latter is well suited to derive brightness profiles
in the case of faint binaries, from which 
the initial values of the model can be
inferred. The least-squares fits are
driven by changes in normalized $\chi^2$, with a noise
model defined for each light curve 
from the data before and after
the occultation.
More details on the instrumentation and the method
can be found in \citet{richichi2011}
and references therein. 
It should be noted that only restricted portions of the light curves
around the main disappearance, corresponding to angular extensions of
$\approx 0\farcs5$, were considered. In general, companions with 
projected separations larger than this would not appear in our list.

\begin{deluxetable}{rllllrrrrrrrl}
\rotate
\tabletypesize{\small}
\tablecaption{List of detected binaries\label{tab:list}}
\tablewidth{0pt}
\tablehead{
\colhead{Seq}&
\colhead{Our ID}&
\colhead{2MASS ID}&
\colhead{Simbad ID}&
\colhead{Date}&\colhead{UT}&
\colhead{B}&\colhead{V}&
\colhead{J}&\colhead{H}&\colhead{K}&
\colhead{Sp.}
}
\startdata
1	&	P85-06	&	07283985+2151062	&	BD+22 1693	&	21-Apr-10	&	00:03:25	&	9.52	&	9.12	&	8.50	&	8.36	&	8.30	&	F2	\\
2	&	P85-23	&	19224512-2046033	&	BD-21 5366	&	21-Aug-10	&	05:39:03	&	11.18	&	10.07	&	8.09	&	7.59	&	7.44	&		\\
3	&	P85-26	&	19240606-2103008	&	BD-21 5373	&	21-Aug-10	&	05:59:25	&	11.63	&	10.28	&	6.73	&	5.93	&	5.65	&		\\
4	&	s033	&	18115684-2330488	&	AKARI-IRC-V1	&	16-Sep-10	&	03:08:18	&	16.50	&	15.60$_{\rm R}$	&	5.70	&	4.51	&	4.02	&		\\
5	&	P86-02	&	22275268-0418586	&	HD 212913	&	15-Nov-10	&	00:52:55	&	10.61	&	8.92	&	5.42	&	4.28	&	4.07	&	M...	\\
6	&	P86-06	&	22282898-0417248	&		&	15-Nov-10	&	01:33:46	&		&		&	10.32	&	9.76	&	9.63	&		\\
7	&	P86-13	&	22304733-0348326	&	HD 213344	&	15-Nov-10	&	03:24:31	&	10.82	&	9.75	&	7.77	&	7.23	&	7.11	&	K0	\\
8	&	P86-21	&	23551978+0532182	&	TYC 593-1067-1	&	17-Nov-10	&	00:35:44	&	10.44	&	9.92	&	8.94	&	8.74	&	8.66	&		\\
9	&	P86-27	&	23581573+0612341	&	TYC 593-1360-1	&	17-Nov-10	&	03:37:07	&	10.85	&	9.88	&	8.44	&	8.04	&	7.95	&		\\
10	&	P86-31	&	23592250+0614580	&	TYC 593-1337-1	&	17-Nov-10	&	04:20:09	&	11.62	&	11.11	&	9.98	&	9.72	&	9.66	&		\\
11	&	P87-014	&	09463209+0913340	&		&	13-Apr-11	&	23:11:43	&		&		&	10.21	&	9.67	&	9.55	&		\\
12	&	P87-026	&	10220366+0515554	&	TYC 252-1217-1	&	11-May-11	&	23:03:59	&	12.03	&	11.33	&	9.86	&	9.48	&	9.42	&		\\
13	&	P87-041	&	10070867+0628201	&	TYC 250-82-1	&	07-Jun-11	&	22:55:04	&	11.23	&	10.55	&	9.53	&	9.27	&	9.15	&		\\
14	&	vmj-023	&	17350423-2319491	&		&	13-Jul-11	&	04:49:05	&		&		&	9.75	&	8.60	&	8.08	&		\\
15	&	vmj-056	&	17395562-2303408	&	HD 160257	&	13-Jul-11	&	07:26:09	&	9.18	&	8.58	&	7.47	&	7.20	&	7.13	&	G2V	\\
16	&	vma-049	&	18142147-2207047	&	IRAS 18113-2208	&	10-Aug-11	&	03:14:03	&		&		&	7.48	&	6.20	&	5.65	&		\\
17	&	vma-055	&	18151114-2213294	&		&	10-Aug-11	&	03:33:19	&		&		&	7.80	&	6.55	&	5.85	&		\\
18	&	vma-062	&	18154557-2220045	&		&	10-Aug-11	&	03:57:43	&		&		&	8.45	&	7.32	&	6.71	&		\\
19	&	vma-083	&	18172695-2147313	&		&	10-Aug-11	&	05:27:34	&		&		&	9.25	&	8.05	&	7.37	&		\\
20	&	P88-003	&	17370333-2239165	&	HD 159700	&	02-Oct-11	&	23:33:40	&	11.28	&	9.67	&	5.81	&	5.04	&	4.60	&	K7	\\
21	&	P88-008	&	17371972-2215045	&	IRAS 17343-2213	&	03-Oct-11	&	00:02:24	&		&		&	8.89	&	7.86	&	7.10	&		\\
22	&	P88-021	&	17401076-2208280	&	IRAS 17371-2207	&	03-Oct-11	&	01:30:13	&	12.16	&	10.67	&	5.32	&	4.11	&	3.75	&		\\
23	&	P88-046	&	18395035-2052114	&	BD-20 5222	&	04-Oct-11	&	02:02:01	&	11.00	&	9.79	&	7.30	&	6.70	&	6.53	&		\\
24	&	P88-050	&	18402840-2040072	&	KO Sgr	&	04-Oct-11	&	02:31:16	&	12.70	&	14.50	&	6.84	&	5.91	&	5.36	&		\\
25	&	P88-052	&	18403433-2034059	&	IRAS 18375-2036	&	04-Oct-11	&	02:45:15	&		&		&	7.89	&	6.85	&	6.31	&		\\
\enddata
\end{deluxetable}

\section{Results}
\label{section:results}
Table~\ref{tab:results} lists our results, following
closely the format already used in previous papers.
The same sequential number used in Table~\ref{tab:list}
is included, followed by the 2MASS identification.
The next two columns are the observed rate of the
event and its deviation from the predicted value.
The difference is due to the local limb slope $\psi$,
from which the actual observed position angle PA
and contact angle CA are derived. We then list
the binary fit results, namely
the signal-to-noise ratio (SNR) of the light curve,
the separation and brightness ratio, and the two
individual magnitudes based on the total magnitude
listed in the 2MASS. As mentioned in Sect.~\ref{section:data},
for Stars 4 and 22 narrow-band filters were used but
we do not expect significant effects on K$_1$, K$_2$.

\begin{deluxetable}{rlrrrrrrccrr}
\rotate
\tabletypesize{\small}
\tablecaption{Parameters of detected binaries\label{tab:results}}
\tablewidth{0pt}
\tablehead{
\colhead{Seq}&
\colhead{2MASS}&\colhead{V (m/ms)}&\colhead{V/V$_{\rm{t}}$--1}&
\colhead{$\psi $($\degr$)}&\colhead{PA($\degr$)}&
\colhead{CA($\degr$)}&\colhead{SNR}&\colhead{Sep. (mas)}&
\colhead{Br. Ratio}&
\colhead{K$_{\rm 1}$}&\colhead{K$_{\rm 2}$}
}
\startdata
1	&	07283985+2151062	&	0.6326	&	-12.0\%	&	-26	&	94	&	-23	&	26.5	&	190.29	$\pm$	0.06	&	2.335	$\pm$	0.004	&	8.69	&	9.61	\\
2	&	19224512-2046033	&	0.4199	&	-8.5\%	&	-4	&	7	&	-58	&	64.2	&	55.5	$\pm$	0.4	&	7.98	$\pm$	0.02	&	7.56	&	9.82	\\
3	&	19240606-2103008	&	0.4715	&	-12.2\%	&	-6	&	107	&	43	&	91.6	&	8.4	$\pm$	0.2	&	12.06	$\pm$	0.04	&	5.74	&	8.44	\\
4	&	18115684-2330488	&	0.6350	&	-5.8\%	&	-5	&	33	&	-38	&	70.3	&	8.8	$\pm$	0.3	&	18.9	$\pm$	0.2	&	4.07	&	7.27	\\
5	&	22275268-0418586	&	0.4576	&	-7.9\%	&	-6	&	83	&	33	&	116.8	&	6.29	$\pm$	0.05	&	7.53	$\pm$	0.01	&	4.20	&	6.39	\\
6	&	22282898-0417248	&	0.2254	&	-21.5\%	&	-6	&	110	&	58	&	10.4	&	6.9	$\pm$	2.6	&	4.5	$\pm$	0.1	&	9.85	&	11.48	\\
7	&	22304733-0348326	&	0.6890	&	10.9\%	&	9	&	283	&	46	&	47.7	&	42.5	$\pm$	0.4	&	29.2	$\pm$	0.3	&	7.15	&	10.81	\\
8	&	23551978+0532182	&	0.5415	&	-7.0\%	&	-10	&	58	&	8	&	39.4	&	15.3	$\pm$	0.7	&	12.7	$\pm$	0.2	&	8.74	&	11.50	\\
9	&	23581573+0612341	&	0.6639	&	0.4\%	&	1	&	47	&	-10	&	46.1	&	19.6	$\pm$	1.0	&	45.1	$\pm$	1.6	&	7.97	&	12.11	\\
10	&	23592250+0614580	&	0.6767	&	-4.2\%	&	-9	&	62	&	2	&	2.5	&	40.1	$\pm$	0.3	&	1.03	$\pm$	0.01	&	10.40	&	10.43	\\
11	&	09463209+0913340	&	0.4553	&	-22.3\%	&	-16	&	141	&	20	&	7.5	&	47.9	$\pm$	0.2	&	1.50	$\pm$	0.01	&	10.11	&	10.54	\\
12	&	10220366+0515554	&	0.6121	&	-3.2\%	&	-4	&	146	&	20	&	9.0	&	29.4	$\pm$	3.5	&	2.34	$\pm$	0.02	&	9.81	&	10.73	\\
13	&	10070867+0628201	&	0.4769	&	5.3\%	&	3	&	179	&	53	&	10.3	&	371.7	$\pm$	4.1	&	1.447	$\pm$	0.007	&	9.72	&	10.12	\\
14	&	17350423-2319491	&	0.3648	&	-0.3\%	&	0	&	137	&	58	&	31.6	&	8.3	$\pm$	0.8	&	9.7	$\pm$	0.1	&	8.19	&	10.66	\\
15	&	17395562-2303408	&	0.6512	&	26.4\%	&	12	&	324	&	67	&	29.3	&	41.96	$\pm$	0.07	&	1.137	$\pm$	0.003	&	7.81	&	7.95	\\
16	&	18142147-2207047	&	0.5564	&	0.1\%	&	0	&	222	&	-32	&	202.4	&	32.4	$\pm$	0.3	&	144.4	$\pm$	1.6	&	5.65	&	11.05	\\
17	&	18151114-2213294	&	0.6585	&	-1.7\%	&	-7	&	71	&	-3	&	209.2	&	15.1	$\pm$	0.2	&	100.1	$\pm$	1.3	&	5.86	&	10.86	\\
18	&	18154557-2220045	&	0.3918	&	-28.0\%	&	-17	&	95	&	22	&	72.9	&	9.9	$\pm$	0.4	&	24.7	$\pm$	0.3	&	6.75	&	10.24	\\
19	&	18172695-2147313	&	0.4303	&	9.8\%	&	3	&	194	&	-58	&	22.5	&	27.8	$\pm$	0.9	&	12.2	$\pm$	0.1	&	7.45	&	10.17	\\
20	&	17370333-2239165	&	0.3548	&	-33.8\%	&	-20	&	98	&	19	&	99.8	&	8.8	$\pm$	0.1	&	25.60	$\pm$	0.09	&	4.64	&	8.16	\\
21	&	17371972-2215045	&	0.4863	&	10.0\%	&	5	&	31	&	-47	&	124.9	&	7.30	$\pm$	0.07	&	18.9	$\pm$	0.1	&	7.16	&	10.35	\\
22	&	17401076-2208280	&	0.6415	&	-1.0\%	&	-1	&	36	&	-40	&	135.5	&	6.2	$\pm$	0.3	&	36.5	$\pm$	0.5	&	3.78	&	7.69	\\
23	&	18395035-2052114	&	0.7922	&	0.4\%	&	1	&	262	&	12	&	109.3	&	748.4	$\pm$	0.2	&	23.13	$\pm$	0.04	&	6.58	&	9.99	\\
24	&	18402840-2040072	&	0.7780	&	-2.1\%	&	-3	&	227	&	-23	&	217.6	&	18.1	$\pm$	0.2	&	78.3	$\pm$	0.6	&	5.37	&	10.11	\\
25	&	18403433-2034059	&	0.5391	&	-15.4\%	&	-9	&	19	&	-52	&	111.9	&	8.1	$\pm$	0.3	&	18.5	$\pm$	0.1	&	6.37	&	9.54	\\
\enddata
\end{deluxetable}

\begin{figure}
\epsscale{0.8}
\plotone{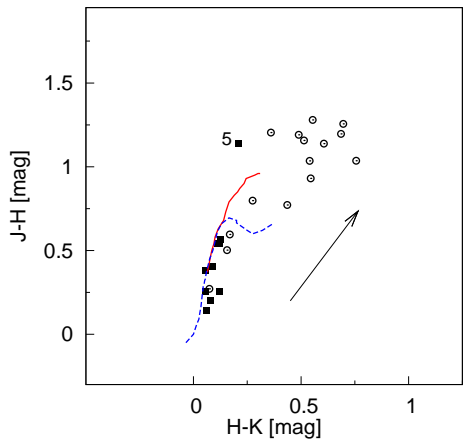}
\caption{2MASS color-color diagram for the sources in our list.
Sources in the direction of the Galactic Bulge are shown
with open circles, and the rest as filled squares.
The lines are the loci of the unreddened giant (solid) and
dwarf (dashed) stars, respectively, according to~\citet{bessellbrett}.
The arrow is the extinction vector for A$_{\rm V}$=5\,mag, according to
~\citet{riekeleb}. The star marked with 5 is discussed
in the text.
}
\label{fig:colors}
\end{figure}

Many of our sources are in the direction of the Galactic
Bulge, having 17$^{\rm h}$ $\la$ RA $\la$ 19$^{\rm h}$ and
Dec $\approx$ $-20\degr$. They have generally very 
red colors; however a color-color diagram shows that
these are mostly consistent with significant amounts
of interstellar extinction (Fig.~\ref{fig:colors}),
with a possible notable exception to be discussed
later. In line with this extinction, many of our sources have
relatively faint optical counterparts, or  none.
Few of them have been  studied in detail previously,
and spectral information is correspondingly scarce.
In the reminder of this section we discuss individual
cases in the context of available literature, when
possible. 

Star 5: 
\object{22275268-0418586} is \object{HD~212913} and \object{IRAS~22252-0434}, 
for which no literature exists except a generic late M spectral type.
The Tycho-2 \citep{2000A&A...357..367H} 
parallax values seem to place this star at $\approx$300\,pc,
thus hinting at a giant star. The star is not in the Bulge and is $49\degr$ below
the galactic disk, yet its location in Fig.~\ref{fig:colors} is peculiar
and indicative of substantial reddening. We attempted fits to our light curve
with a binary, a resolved diameter, and a diameter plus companion models.
The first one gave the best normalized $\chi^2$ and is the one we adopt in our results.


Star 12:
\object{10220366+0515554} was included in Tycho-2 
 as~\object{TYC~252-1217-1}. However, no binarity was detected, including in the
subsequent dedicated re-analysis of the Tycho Double Star 
 Catalogue~\citep{2002A&A...384..180F}. 
 No literature references were found for this source.

Star 13:
\object{10070867+0628201} is
\object{TYC~250-82-1}. In spite of the relatively  large measured projected separation,
 no binarity was detected in Tycho 2.
 No literature references are present for this source.





\begin{figure}
\includegraphics[angle=-90.0, width=7.5cm]{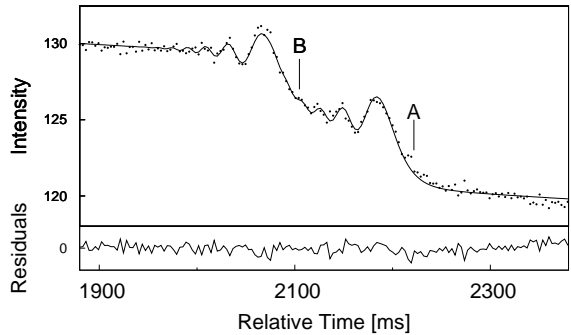}
\caption{Occultation light curve (dots) and
best fit (solid line) for HD~160257, a newly detected
pre-main sequence
binary. The times of
geometrical occultation of the two components
are marked. The lower panel shows the fit residuals.
}
\label{fig:vmj-056}
\end{figure}

Star 15:
\object{17395562-2303408} is \object{HD~160257}, a bright, nearby, G2V-type star.
It was classified as a 
pre-main sequence 
star by \citet{2006A&A...460..695T}, on the basis of
X-Ray emission and 
high resolution spectra. We find it to 
to be
binary with a rather large projected separation
(see Fig.~\ref{fig:vmj-056}). It was observed by Hipparcos as 
\object{HIP 86455}, but no binarity was found.
As for the recent case of new binaries in the Pleiades
\citep{2012A&A...541A..96R}, this
shows the potential of LO to further extend the statistics of
binarity among young stars in the context of multiple star
formation mechanisms.

Star 16:
\object{18142147-2207047} is the infrared source 
\object{IRAS 18113-2208}. No literature references were found for this source.
It is also the binary with the largest magnitude difference in our sample,
 $\Delta$K=5.4\,mag against a dynamic range of 5.8\,mag.



\begin{figure}
\includegraphics[angle=-90.0, width=7.5cm]{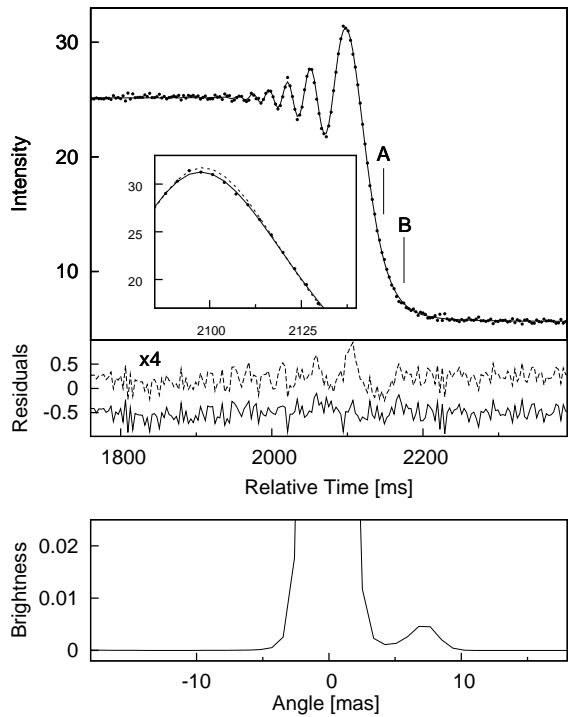}
\caption{Top: occultation data (dots) for
P88-08 = IRAS~17343-2213, and best fit by a binary
source model (solid line). The residuals for the best fit by a
single source ($\chi^2_n=1.71$) and a binary source model
($\chi^2_n=1.13$) are
shown, enlarged for convenience, in the lower panel as
dashed and solid lines, respectively.
The inset shows an enlargement of the single source (dashed line)
and binary source (solid line) models around the first fringe.
Bottom: a portion of the brightness profile recovered by the
CAL model-independent method.
}
\label{fig:p88-08}
\end{figure}

Star 21:
\object{17371972-2215045} is the infrared source
\object{IRAS 17343-2213}, also measured by the AKARI satellite.
There are no bibliographical references. It is one of the reddest
sources in our sample,
having J-K=1.8\,mag from 2MASS. Our recorded flux shows
an increase of $\approx$80\% above the 2MASS K$_{\rm s}$ magnitude
(cfr. Fig.~\ref{fig:counts}),
pointing to possible variability. The detection of the 3.2\,mag fainter
companion with 7\,mas projected separation is shown in Fig.~\ref{fig:p88-08}.
In the lower panel of the figure, note that the CAL algorithm
preserves the integrated flux ratios, not the peaks in the 
brightness profile.

Star 23: \object{18395035-2052114} is the star
with the largest projected separation in our list,
$0\farcs75$, however it does not seem to have
been detected before in spite of being
relatively bright. The 1:23 brightness ratio in the K-band,
and possibly more at visual wavelengths, could be
one reason. Concerning projection effects of
a nearby field star,
we have investigated  images from DSS, HST and 2MASS
without evidence of other stars within the ISAAC
subwindow.

Star 24:
\object{KO Sgr}, also identified as \object{IRAS 18375-2042}, was found to
have a regular period of 312~days by \citet{Hof60} and thus
tentatively classified as a Mira-type star. The same author, however,
noted the peculiarity of rather steep increases in luminosity,
and the relatively short-lived maxima. She suggested the possibility
of a companion in a cataclysmic system, or just a visual association. 
In this scenario, the light at minimum would come mostly from the secondary.
The photographic magnitude difference between minima and maxima is
about 3.1\,mag. Using the Hoffleit photometric period, our observation
would have occurred at phase 0.8 or just before the onset of the
outburst-like maximum. Uncertainties are possible due to the $\approx$100
cycles intervened since Hoffleit's first determination, but are not
easy to estimate. Data in the AAVSO database are also not very complete.
Under the above assumptions,
the $\Delta$K=4.7\,mag would point
to a primary much redder than the secondary, consistent with the scenario
outlined in \citet{Hof60}.

We also recorded a LO light curve of 
\object{08550318+1346301} = \object{BD+14 1995} on April 13, 2011 (star P87-010 in our
database), detecting a well-separated companion with $385.5 \pm 0.9$\,mas
projected separation along PA=$161\degr$. The 2MASS-based magnitudes of
the two components are K=9.04 and K=10.31.
This is listed as the wide double 
\object{WDS J08551+1347} \citep{2001AJ....122.3466M}, without orbital parameters
and with a separation of $1\farcs4$. Given the wide nature of this binary,
we did not include it in our tables and figures, and we mention our
results only as a complement to the previous observations.

\section{Discussion and Conclusions}
\label{sec:conclusions}
\citet{apjssur} list the sources
observed in the same program and found to be unresolved, 
covering also the period under consideration here.
From the first to the last
night considered in Table~\ref{tab:list},
a total of 403 LO light curves were observed at the
VLT with ISAAC, therefore our serendipitous binary
detection fraction is (26/403) $\approx 6.5$\%. Restricting ourselves
to the purely random service mode observations, the
number is (19/231) $\approx 8.2$\%.
We note that the
visitor mode targets were in crowded and extincted regions in,
and consequently mostly deep into,
the Galactic Bulge, with a median K=6.4\,mag.
By contrast, the service mode targets were randomly scattered
and with  median K=8.0\,mag. One possible reason for the
higher incidence of binaries among the service mode targets
is that they are statistically closer to us, therefore providing
a better spatial scale for the same angular resolution.

\begin{figure}
\includegraphics[width=7.5cm]{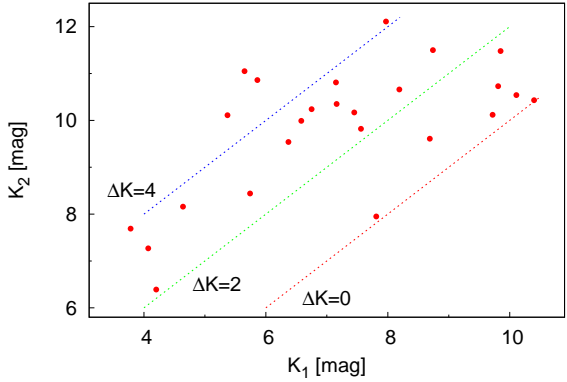}
\caption{Distribution of the K magnitudes of the primaries
and secondaries in our sample, as computed from the
flux ratio in our light curves and the 2MASS K$_{\rm s}$
values. 
}
\label{fig:k1k2}
\end{figure}

In a previous work based on a comparable volume of
LO observed also in the near-IR
in subarray fast mode 
from Calar Alto \citep{2006A&A...445.1081R}, it was found that the
serendipitous binary detection frequency was
significantly smaller than the present one, $\approx 4$\%.
The difference can be justified in principle with the lower sensitivity
(limiting K $\approx 8$\,mag then, and $\approx 12$\,mag here)
and the slower time sampling ($\approx 8.5$\,ms then, against
3.2\,ms here). However, both cases appear to have an inferior
yield of binary stars than observations by a fast photometer:
e.g., \citet{richichi96} using the TIRGO 1.5\,m telescope
found (26/157) $\approx 16$\%
binary detection frequency. The explanation in this case is 
that the targets were brighter, being selected prior to the
availability of the 2MASS catalog, with average K $\approx 4$\,mag.
Hence, the sources were generally significantly closer to us
than in the present study.

\begin{figure}
\includegraphics[width=7.5cm]{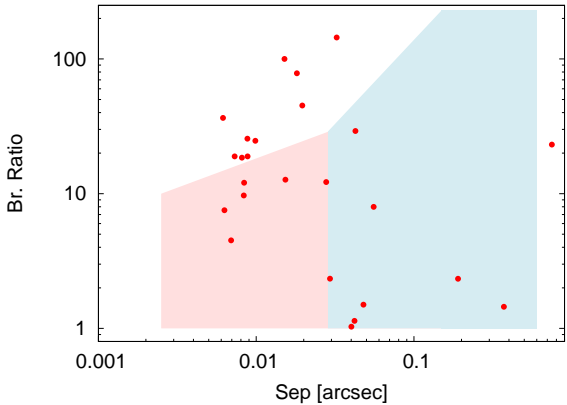}
\caption{Distribution of brightness ratios against
projected separations for the stars in our sample.
The shaded areas mark the approximate
regions where near-infrared long-baseline
interferometry (left) and adaptive optics at an 8\,m
telescope (right) are mostly effective.
}
\label{fig:brsep}
\end{figure}

The flux ratios of the binaries in our list range
from $\Delta$K=0.0 to 5.4\,mag, with a median of 2.8\,mag.
From Fig.~\ref{fig:k1k2} it can be seen that half of
the sample falls in the range 2$\le \Delta{\rm K} \le$4\,mag.
As for the separations, LO can measure only projected values,
and the median for our sample is 18\,mas. 
Considering a random projection correction of 2/$\pi$, about half of our binaries
would remain inaccessible to techniques with less than $0\farcs03$
resolution. This corresponds, e.g., to the K-band diffraction limit
of a 12\,m telescope. Long-baseline interferometry at some of the
largest facilities  can provide
the necessary angular resolution, but it suffers from a sensitivity
which is significantly more limited than for LO and in general from a reduced
dynamic range. We outline schematically the situation in Fig.~\ref{fig:brsep}.
From the figure, it can be estimated that only about 2/3 of the
sources in our list might be effectively followed up by independent methods.

In conclusion, we find from the present work as well as from
previous similar research that a routine program of 
random LO with an IR detector
operated in subarray mode at an 8\,m-class telescope
with time resolution of order 3\,ms
can detect companions to stars in the general field
down to a sensitivity K $\approx 12$\,mag, with separations
as small as a few milliarcseconds (at least 10 times better
than the diffraction limit of the telescope).
The
expected detection frequency can range from $\approx 6$\%
in distant regions such as deep in the Galactic Bulge, 
to $\approx 15$\% in areas closer to the solar neighbourhood.

Most of these binaries or multiples turn out to be new
detections, with limited cross-identifications, spectral
determinations and previous literature.
Due to the small separations and high brightness ratios,
only up to $\approx 50$\% of these LO-detected binaries can be
followed up by other more flexible methods such as 
adaptive optics and interferometry. A significant fraction
of such new binary detections will remain isolated until 
significantly larger telescopes or more sensitive interferometers
become available.

\acknowledgments
We are grateful to the ESO staff in Europe and Chile for
their continued support especially of the
service observations.
OF acknowledges financial support from MINECO through
a {\it Juan de la Cierva} fellowship and from \emph{MCYT-SEPCYT Plan Nacional I+D+I AYA\#2008-01225}.
This research made use of the Simbad database,
operated at the CDS, Strasbourg, France, and
of data products from the Two Micron All Sky Survey (2MASS), 
which is a joint project of the University of Massachusetts 
and the Infrared Processing and Analysis Center/California Institute 
of Technology, funded by the National Aeronautics and 
Space Administration and the National Science Foundation.
This research has made use of the Washington Double Star Catalog
maintained at the U.S. Naval Observatory.

\clearpage

\end{document}